\documentclass[11pt,a4paper]{article}
\usepackage{jcappub}

\usepackage{color}

\title{The behaviour of shape and velocity anisotropy in dark matter haloes}
\author[]{Martin Sparre}
\author[]{and Steen H. Hansen}
\affiliation[]{Dark Cosmology Centre, Niels Bohr Institute,\\
University of Copenhagen, Juliane Maries Vej 30, 2100 Copenhagen, Denmark}

\emailAdd{sparre@dark-cosmology.dk}
\emailAdd{hansen@dark-cosmology.dk}

\abstract{
Dark matter haloes from cosmological N-body simulations typically have triaxial shapes and anisotropic velocity distributions. Recently it has been shown that the velocity anisotropy, $\beta$, of cosmological haloes and major merger remnants depends on direction in such a way that $\beta$ is largest along the major axis and smallest along the minor axis. In this work we use a wide range of non-cosmological N-body simulations to examine halo shapes and direction-dependence of velocity anisotropy profiles. For each of our simulated haloes we define 48 cones pointing in different directions, and from the particles inside each cone we compute velocity anisotropy profiles. We find that elongated haloes can have very distinct velocity anisotropies. We group the behaviour of haloes into three different categories, that range from spherically symmetric profiles to a much more complex behaviour, where significant differences are found for $\beta$ along the major and minor axes.
We encourage future studies of velocity anisotropies in haloes from cosmological simulations to calculate $\beta$-profiles in cones, since it reveals information, which is hidden from a spherically averaged profile.
Finally, we show that spherically averaged profiles often obey a linear relation between $\beta$ and the logarithmic density slope in the inner parts of haloes, but this relation is not necessarily obeyed, when properties are calculated in cones.
}


\begin{document}
\maketitle

\section{Introduction}

The gravitational potential of galaxies and galaxy clusters are dominated by dark matter haloes. Such haloes have only been observed through their gravitational impact on other objects: e.g. in the Bullet Cluster \citep{2006ApJ...648L.109C,2006ApJ...652..937B}, the motion of galaxies in clusters \citep{1933AcHPh...6..110Z} and in the cosmic microwave background \citep{2011ApJS..192...18K}. Therefore, the most fruitful way to study the detailed dynamics of dark matter structures is through numerical simulations.

The formation and evolution of the large-scale structure in the universe can be simulated with N-body codes, where the density field and velocity field of the dark matter are represented by collisionless particles. In such simulations it is found, that the dark matter haloes follow a universal density profile, usually parametrised as an Einasto-profile \citep{2004MNRAS.349.1039N} or a NFW-profile \citep{1996ApJ...462..563N}. A different universality is the pseudo-phase-space density profile, $\rho / \sigma_\text{rad}^3$, which is a power law in radius ($\rho$ is the density and $\sigma_\text{rad}$ is the radial velocity dispersion) \citep{2001ApJ...563..483T,2004MNRAS.352L..41H,2005ApJ...634..756A,2005MNRAS.363.1057D}. It is worth noting that particles in subhaloes and streams do not follow this power law behaviour \citep{2011MNRAS.tmp..937L}. Subhaloes are also responsible for large peaks in the true 6-dimensional phase-space density distributions in haloes from cosmological simulations \citep{2004MNRAS.353...15A,2005MNRAS.356..872A, 2006MNRAS.373.1293S,2009MNRAS.398L..21S,2010MNRAS.406..137L}.

The mass distributions of haloes from cosmological simulations are triaxial ellipsoids with a major axis (with length $a$), an intermediate axis (with length $b$) and a minor axis (with length $c$). If $a\simeq b > c$ a halo is oblate (\emph{pancake} shaped), if $a > b \simeq c$ a halo is prolate (\emph{cigar} shaped), and if $a\simeq b\simeq c$ a halo is spherical. High-redshift haloes are typically more elongated than those at low redshift, and it is also a trend that light haloes are closer to being spherical than massive haloes \citep{2006MNRAS.367.1781A}. One possible mechanism that likely is important for creating elongated haloes is mergers, that can produce both prolate and oblate haloes \citep{1978MNRAS.184..185W,2004MNRAS.354..522M}.

Dark matter systems can have anisotropic velocity distributions. The anisotropy is characterised by the velocity anisotropy parameter,
\begin{align}
\beta(r) \equiv 1 - \frac{\sigma_\text{tan}^2}{2\sigma_\text{rad}^2},
\end{align}
where $\sigma_\text{rad}$ is the velocity dispersion in the radial direction, and $\sigma_\text{tan}$ is the total tangential velocity dispersion. $\beta$ is positive in regions with radially dominated anisotropy, and negative in regions with tangentially dominated anisotropy. The inner parts of haloes in cosmological simulations have $\beta$-profiles (calculated in spherically averaged bins) increasing from 0 in the deep interior to about 0.3 at the core radius \citep{2004MNRAS.352..535D,2008MNRAS.389..385C,2011MNRAS.tmp..937L,2012ApJ...752..141L}. The $\beta$-profiles in the outer parts exhibit different behaviour from halo to halo.

From the outcome of a wide range of simulations \citep{2006NewA...11..333H}, a relation has been suggested between the spherically averaged value of $\beta$ at a given radius ($r$) and the slope of the density profile,
\begin{align}
\gamma (r)\equiv \frac{d\log \rho}{ d\log r}.
\end{align}
The relation, $\beta (\gamma)=-0.2\times( \gamma + 0.8)$, is valid in the inner parts of haloes in cosmological simulations \citep{2006JCAP...05..014H}. A more recent relation, the \emph{attractor}, was found in several different structures, which were exposed to an artificial process that instantaneously interchanged energy between particles \citep{2010ApJ...718L..68H} (HJS). The role of the attractor in cosmological simulations, however, remains undetermined.

There exists evidence that $\beta$-profiles are not spherically symmetric functions (or functions that are constant along density or potential contours, if the structure is non-spherical) in all dark matter haloes. In a detailed study of remnants of collisionless mergers, it has been shown that the velocity anisotropy profiles of merger remnants exhibit different behaviour depending on whether they are calculated along the major or the minor axis \citep{2012arXiv1205.1799S}. The behaviour of $\beta$ were clearly correlated with the shape and the orientation of the haloes, in such a way that the largest velocity anisotropy was found along the major axis. Similar asymmetric $\beta$-profiles are also found in cosmological simulations \citep{2009MNRAS.394..641Z}.

The aim of this article is to study the behaviour of direction-dependent velocity anisotropy profiles and halo shapes in a set of non-cosmological simulations. In section~\ref{ch:sim} we describe our simulations, and in section~\ref{ch:analysis} we describe our analysis methods. Next (section~\ref{ch:betashapes}) we study the direction-dependence of $\beta$-profiles together with halo shapes, and we compare our results to cosmological haloes in section~\ref{ch:cosmo}. In section~\ref{ch:sphaverage} we report halo shapes and spherically averaged $\beta$- and $\gamma$-profiles. Finally we compare haloes with the attractor from HJS (section~\ref{sec:attractor}).

\section{Simulations}\label{ch:sim}

The overall purpose of our simulations is to expose haloes to a range of perturbations, so we can study velocity anisotropy profiles and halo shapes in haloes formed in several different ways. Simulation I is an artificial simulation, that involves an instantaneous change in the gravitational potential, simulation II is the energy exchange perturbation from HJS, simulation III is a collapse experiment, simulation IV involves multiple major mergers, simulation V involves unstable haloes and simulation VI shows the effect of substructure in a halo. An overview of parameters is given in Table~\ref{Table:Sim}, and an overview of the initial structures is given in Table~\ref{Table:structures}.

\subsection{Simulation code}

To run our simulations we used the N-body simulation code, Gadget-2 \citep{2001NewA....6...79S,2005MNRAS.364.1105S}. The simulations were run in a non-cosmological Newtonian box, and we only used collisionless particles. For all the simulations the spline softening, $\eta$, implemented in Gadget-2 was used. The time-step of each particle was calculated as $\Delta t = (2 \eta \epsilon / |\textbf{a}|)^{1/2}$ (for a discussion of time-step criteria see \citep{2007MNRAS.376..273Z}), where $|\textbf{a}|$ is the magnitude of the acceleration and $\epsilon$ is the accuracy parameter, which was set to 0.05 in our simulations.  In all simulations the energy conservation was better than 1.0 $\%$.

\begin{table}
\centering
\caption{The softening, the total mass and the gravitational constant in the simulations. $r_\text{s}\equiv 1$ in all the initial structures.}
\label{Table:Sim}
\begin{tabular}{cllrc} 
\hline\hline 
 & Description & Softening & Mass & $G$   \\\hline
I  & $G$-variations    & 0.0050 &1 &$0.8-1.2$    \\
II  & Energy exchange (HJS) & 0.0050&1 & 1    \\
III & Cold collapse & 0.0025 & 1 & 1 \\
IV & Multiple mergers& 0.0050 &128 & 1\\
V & Unstable models & 0.0050 &1 & 1  \\
VI & Substructure &0.0015&1 & 1\\
\hline\hline
\end{tabular}
\end{table}

\begin{table}
\centering
\caption{An overview of the structures in the simulations. The \emph{Colour} and \emph{Symbol} columns refer to how structures are presented in figure \ref{Fig01:betasph}-\ref{gamma_kappa}.\newline$^\text{a}$ Same collision axis in the two last simulations.}\label{structuretabel}
\label{Table:structures}
\begin{tabular}{lllllllll} 
\hline\hline 
Name&$\rho$ & $\beta$ & Sim.  & Colour & Symbol \\\hline
A&Hernquist & 0 & I and II & maroon  &$\Diamond$ \\
B&$1/(1+r)^5$ & $0$ & I and II  & red &$\circ$  \\
C&Hernquist & $r^2/(1+r^2)$, $r_\text{an}=1.0$& I and II  & orange &$\triangleleft$ \\
D&$1/(1+r)^{3.5}$ & $r^2/(1.5^2+r^2)$, $r_\text{an}=1.5$ & I and II  & blue &$x$  \\
E&Hernquist & $0$, Gaussian & I and II & pink &$*$ \\ 
F&Hernquist & $1/2$, Gaussian & I and II & green &$+$ \\
G&Hernquist &  $r^2/(1+r^2)$, Gaussian & I and II & black &$\triangleright$  \\
H&Hernquist &  $1-r^2/(1+r^2)$, Gaussian & I and II & yellow &$\triangledown$ \\ 
\hline
I&Hernquist & - & III  & black &$\circ$  \\
J&$1/(1+r)^5$ & - & III  & black &$x$  \\
\hline
K&Hernquist & $0$ & IV  & red &$\Diamond$  \\
L&$1/(1+r)^5$ & $0$  & IV & red &$*$  \\
M&$1/(1+r)^{3}$ & $0$ & IV  & red &$\triangleleft$ \\
N&Hernquist & $0$ & IV$^\text{a}$  & blue &$\circ$ \\
O&$1/(1+r)^5$ & $0$  & IV$^\text{a}$  & blue&$*$  \\
P&$1/(1+r)^{3}$ & $0$ & IV$^\text{a}$  & blue &$\triangleleft$  \\
\hline
Q&Hernquist & $r^2/(0.2^2+r^2)$, $r_\text{an}=0.2$ & V   & green, IC: blue &   $\triangleleft$\\
R&$1/(1+r)^5$ & $r^2/(0.2^2+r^2)$, $r_\text{an}=0.2$ & V   & green, IC: blue &   $*$ \\
\hline
S&Hernquist & $0$ & VI   & red, IC: black &   $\Diamond$, IC: $\circ$\\
\hline\hline
\end{tabular}

\end{table}

\subsection{Simulation I -- An instantaneous change in the potential} \label{SimIIII}

The first of our simulations will show how structures respond to a process, where the gravitational potential changes instantaneously. Such a change perturbs the accelerations in contrast to the perturbations of the velocities performed in HJS (see section~\ref{HJSChapter} for details).

First eight structures with different velocity anisotropies and density profiles were created. The density profiles were of the form,
\begin{align}
\rho(r) = \frac{\rho_0}{\left(r/r_\text{s}\right)^{-\xi}}\frac{1}{\left(1+r/r_\text{s}\right)^{-\zeta}},
\end{align}
where $r_\text{s}$ is the scale radius and $\rho_0$ is a normalization constant determining the total mass.
Six structures followed a Hernquist density profile ($\xi=1$ and $\zeta=3$) \citep{1990ApJ...356..359H}, one structure had $(\xi,\zeta)=(0,3.5)$ and one structure had $(\xi,\zeta)=(0,5)$.

Two of the structures were created using Eddington's formula for $\beta=0$ \citep{2008gady.book.....B}, and two structures were Osipkov-Merritt models \citep{1985AJ.....90.1027M,1979SvAL....5...42O} with velocity anisotropy profiles given by,
\begin{align}
\beta (r) = \frac{r^2}{r_\text{an}^2+r^2},
\end{align}
where $r_\text{an}$ is the anisotropy radius.

For four of the Hernquist structures the initial velocity distributions were Gaussian distributions with velocity dispersions, $\sigma_\text{rad}$ and $\sigma_\text{tan}$, calculated from the Jeans equation \citep{1993ApJS...86..389H}. Structures with $\beta_\text{initial} = 0$, $\beta_\text{initial} = 1/2$, $\beta_\text{initial} = r^2/(1+r^2)$ and $\beta_\text{initial} = 1-r^2/(1+r^2)$ were created. Structures generated with this method are not in perfect equilibrium, so simulations were run for 100 time units, where the structures had time to equilibrate. We also ran such test simulations for the structures generated with Eddington's method and for the Osipkov-Merritt models to assure equilibration.

All structures had $r_\text{s}=1$, and $\rho_0$ was chosen such that the total mass inside $r=13r_\text{s}$ was 1. When structures are created in this way the dynamical time, $1/\sqrt{G\overline{\rho}}$, for particles inside $r=13r_\text{s}$ is smaller than 100 time units, which is the duration of all our simulations. The initial conditions were generated with $G=1$. The structures used in this simulation are identical to those used in HJS. The initial density profiles and velocity anisotropy profiles are summarised in table~\ref{Table:structures}.

The next step in our setup was to run a simulation for 100 time units, where the gravitational constant was set to $G=0.8$. In this time the structures expanded due to the lowering of the potential. Next the gravitational potential was increased by setting $G=1.2$, which caused a contraction in the following 100 time units. We kept changing between $G=1.2$ and $G=0.8$ every 100 time units until 2000 time units had passed. In the last of these simulation we had $G=1.2$. An additional simulation with $G=1.2$ was run for 100 time units, so the structures had more time equilibrate.

\subsection{Simulation II -- Exchange of energy}\label{HJSChapter}

Simulation II is the simulation from HJS, where an exchange of kinetic energy between particles in the same spherical bin was performed. The experiment was performed for the haloes described in section~\ref{SimIIII}. In each perturbation the kinetic energy of each particle was multiplied by a uniformly chosen random number in the range $[0.25,1.75]$. Conservation of the total energy was taken care of by scaling all the kinetic energies by a constant (the constant was typically between 0.98 and 1.02). After this perturbation the haloes were evolved in time with a N-body simulation code. 20 of such perturbations (followed by time evolution of the system) were performed. See more details in HJS.

\subsection{Simulation III -- A cold collapse experiment with substructure}

In this simulation we want to assemble a structure through a violent relaxation process (similar to \citep{1982MNRAS.201..939V}), that mimics the way structures are assembled when they collapse in the early universe. Our setup will be a main halo, which contains several compact substructures. Initially, all particles had a velocity of zero, so this simulations is therefore effectively a collapse simulation, where the substructures break the spherical symmetry.

We distributed $5\times 10^5$ particles according to a Hernquist density profile with $r_\text{s}=1$ and a cutoff at $200r_\text{s}$. Next $5\times 10^5$ particles, with the same total mass as the main halo particles, were distributed in 24 identical subhaloes, that also followed Hernquist profiles, but with a scale radius of 0.5 and a cutoff radius of 5. The positions of the subhaloes were sampled in the same way as the particles in the main halo. The total mass in the simulation was 1. With this setup the time scale of the collapse is 2.45 times larger for the main halo than in the subhaloes.

We ran the simulation for 200 time units, which corresponds to 200
dynamical times at the scale radius for the initial structure. An additional simulation was run, where the initial density profiles of the main halo and the subhaloes were,
\begin{align}
\rho(r) = \frac{\rho_0}{\left(1+r/r_\text{s}\right)^5}. \label{rho05}
\end{align}
We defined $r_\text{s}=1$, and $\rho_0$ was found by defining the total mass of the halo to be 1.

Note that this simulation is a physically realistic experiment, in the sense that no artificial perturbations has entered the simulation, even though the initial condition for the collapse is different from that of cosmological simulations, e.g. because the initial conditions have a cuspy density profile.

\subsection{Simulation IV -- Major mergers}\label{MajorMergers}

Now we will build up a structure through major mergers. We first generated two identical structures with Eddington's formula with $\beta=0$. The two structures were collided, and the remnant of this
collision was duplicated and collided again. This procedure was repeated until seven
collisions were done. In the collisions the impact parameters were 0, and the collision axis changed
from simulation to simulation\footnote{The collision axes were the following in the seven simulations: $x$, $y$, $z$, $x$, $y$, $z$, $x$. The structures were not rotated between the simulations.}. The initial distances between the collided structures were $10.0$,
$12.5$, $15.0$, $25.0$, $35.0$, $40.0$ and $50.0$, respectively, and the structures had initial
relative velocities of 0.

There were $10^4$ dark matter particles in the first simulation and $1.28\cdot 10^6$ in the last simulation. We used units where the total mass and $r_\text{s}$ of the initial structure were defined to be $1$, and a cutoff was made at a radius of $10$. We ran the last simulation for 200 time units, so the structures had time to first collide (it happened after $\sim75$ time units) and then form a new equilibrated halo.

The simulations were done for initial structures following three different density profiles: a
Hernquist profile,  $\rho(r)\propto 1/(1+r)^3$ and Eq.~\eqref{rho05}.

Additional simulations were run with the same three density profiles, but instead of rotating the collision axis in the last simulation we used the same collision axis in simulation 6 and 7\footnote{So the collisions occurred along the following axes in the mergers: $x$, $y$, $z$, $x$, $y$, $z$, $z$.}.

\subsection{Simulation V -- Unstable Osipkov-Merritt models}

It is established that Osipkov-Merritt models following Hernquist profiles are unstable for low values of $r_\text{an}$ \citep{1997ApJ...490..136M}, due to the onset of a radial orbit instability \citep{1985MNRAS.215..517B}, which creates a bar structure in the center of the halo. In this simulation we will study how such an instability affects the velocity anisotropy of a system. The perturbations in this simulation are therefore non-spherical because of the formation of a bar, despite the fact that initial conditions are spherically symmetric.

We ran simulations with two density profiles; a Hernquist profile and Eq. \eqref{rho05}. In both cases we defined $r_\text{s}=1$ and $r_\text{an}=0.2r_\text{s}$. The total mass was 1. The two simulations were run for 200 time units and $10^6$ collisionless particles were used to represent each halo.

\subsection{Simulation VI -- A halo and its subhaloes}\label{sub}

To see how cosmologically realistic haloes are affected by the dynamical friction from the subhaloes
inside them, we generated a halo and subhaloes similar to what is found in the Via Lactea II simulation, which contains a main halo with a mass of $1.94\times 10^{12} M_\odot$ and a tidal radius of 462 kpc \citep{2008Natur.454..735D}.

We used $10^6$ dark matter particles to represent the halo and the subhaloes, but only subhaloes heavier than $1.94\times 10^8 M_\odot$ (the mass of 100 particles) were included. In total 7\% of the mass was contained in subhaloes. All haloes followed Hernquist density profiles with the same radius of maximum circular velocity, tidal radius, tidal mass, position and velocity as in the public catalog from the Via Lactea II simulation. The velocities of the main halo particles and the subhalo particles were sampled from velocity distributions calculated using Eddington's formula with $\beta=0$. We ran the simulation for 100 time units, which is the same as 100 dynamical times for a particle at the scale radius.

\section{Analysis methods}\label{ch:analysis}

\subsection{Analysing particles in cones}
In our analysis we will study particles in cones pointing in different directions. Each cone has an apex angles of $45^\circ$. For each particle and axis we calculate $\theta = \arccos (\hat{\mathbf{n}}\cdot \mathbf{r} / |\mathbf{r}|)$, where $\hat{\mathbf{n}}$ is a unit vector pointing in the direction of a cone, and $\mathbf{r}$ is a particle's position vector (seen from the center of the halo). Particles are selected to be contained in a cone, if $\theta\leq 22.5^\circ$. Analysing structures in this way was also done in \citep{2012arXiv1205.1799S}.

We created 48 cones centered on the most bound particle of each halo. To define the pointing direction of each cone we generated 48 points on a sphere.  The points were distributed in the same way as a sphere is divided into pixels in typical studies of the cosmic microwave background \citep{2005ApJ...622..759G}. Note, that some cones will be overlapping each other in space, because of our choice of apex angle and number of cones.

\subsection{Determination of halo shapes}\label{shapedetermination}

Halo shapes and orientations are determined from the eigenvectors and eigenvalues of the moment of inertia tensor. First the shape in the central part is fitted with an ellipsoid, and next the shape is determined as function of radius with the procedure described in detail in \citep{1991ApJ...368..325K}. In the remaining part of this article we will refer to the shape found with this method as the \emph{shape} of a halo. In the following section we will also discuss contours of constant $\rho$ and $\beta$ in haloes, and in these cases we will explicitly mention, what kind of contour we are working with, so the different types of contours can be clearly distinguished.

Note that we in this article let $r$ denote the physical radius, and not the elliptical radius as it is done in some other studies.

\section{The behaviour of velocity anisotropy and halo shape} \label{ch:betashapes}

To analyse the behaviour of shape and the direction-dependence of the velocity anisotropy for our haloes, we will present plots of $\beta(r)$ and $\rho(r)$ through cones for a representative selection of haloes from our simulations.

\subsection{Category 1: Haloes with well-behaved $\beta$-profiles}

Figure \ref{s4GBetaCones} shows halo H from simulation II, where structures were perturbed by varying $G$ in time. Each of the lines represent a property ($\beta$ or $\rho$) calculated through a cone for the structure. Also shown (in the right panel) are the axis ratios, $c/a$ and $b/a$, calculated from the inertia tensor. The plotted radii are in units of $r_{-2}\equiv r(\gamma=-2)$.

The right panel shows the prolate shape of the halo. From the panel that shows $\rho$ in different cones, it can also be seen that the halo is elongated since there is an offset between the densities in different cones. From the plot of $\beta$ in different cones, it is seen that $\beta$ is systematically larger along the major axis than along the minor axis at any given radius.

The contours of constant $\beta$ are more elongated than the contours of constant $\rho$. The minor to major axis ratio of each contour can be approximated by the length of the cyan line segment in figure~\ref{s4GBetaCones}, and it is found that the $\beta$-contour has a minor to major axis ratio of 0.1, and that the $\rho$-contour has a minor to major axis ratio of 0.6.

Halo A from simulation II, where a sudden change of particle energies was made, is shown in figure~\ref{0EBetaCones}. This halo is also close to being prolate, and we again see that the largest velocity anisotropy is in the direction of the major axis\footnote{When 48 cones with apex angles of $45^\circ$ are distributed on a sphere, it can happen that the minor axis (or one of the other principal axes) is contained in several cones, since some of the cones are overlapping. This is why several cones are marked as being along the minor axis in Figure~\ref{0EBetaCones}.}. The minor to major axis ratio of the $\beta$-contour is 0.35 and the axis ratio from the density shape is 0.49, so we again see that the $\beta$-contour is more elongated than the $\rho$-contour.

\subsection{Category 2: Haloes with complex $\beta$-profiles}

For halo H from the collapse simulation (figure~\ref{ColdCollapseBetaCones}) a slightly more complicated behaviour of $\beta$ is seen. In the inner parts $\beta$ is smallest along the minor axis of the density ellipsoid, and in the outer parts it is largest along this axis. Such a \emph{crossover} is absent in the plot of $\rho$ through different cones. Because of the crossover, the $\beta$-contours of the halo do not have an ellipsoidal shape.

Figure \ref{MergerCones} and \ref{ROI2} show more examples of haloes with $\beta$-profiles that are clearly not ellipsoidal. For the merger remnant $\beta$ is roughly constant and positive along the major axis and monotonically increasing along the minor axis. This behaviour is consistent with our previous study of major mergers \citep{2012arXiv1205.1799S}. The instability simulation gives a complicated behaviour of both $\beta$, $\rho$ and the axis ratios. Is is e.g. seen that $\rho$ has significant wiggles and bumps along the minor axis.

\subsection{Category 3: The spherically symmetric haloes}

Finally we will also note that perturbed haloes also can follow simple spherically symmetric distribution functions. Figure~\ref{0GBetaCones} shows halo A from simulation I, and it has a spherical shape, and $\beta$-profiles behaving similarly from cone to cone.

\begin{figure*}
\centering
\includegraphics[width=\linewidth]{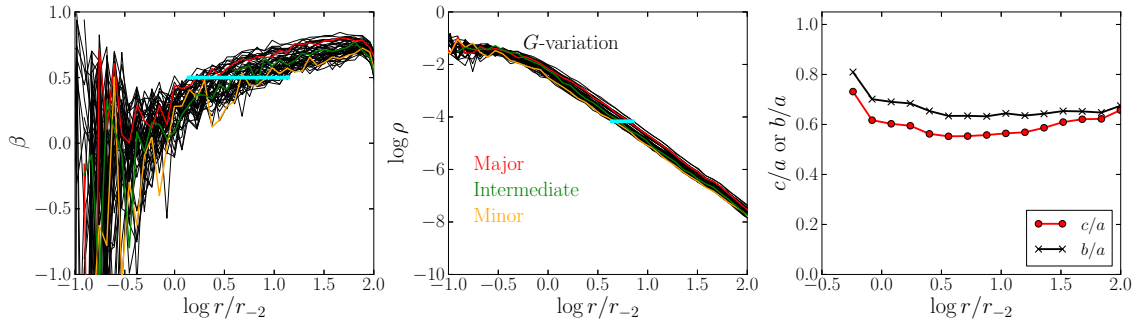}
\caption{Velocity anisotropy profiles (\emph{left panel}) and densities (\emph{central panel}) calculated in 48 cones for halo H (a Hernquist profile with $\beta_\text{initial}=1-r^2/(1+r^2)$) from the simulation, where $G$ was changed instantaneously. Each line corresponds to a cone, and the coloured lines corresponds to cones which coincides with one of the principal axes. The \emph{right panel} shows the axis ratios as function of radius. The cyan horizontal line segment in the \emph{left} (\emph{central}) \emph{panel} shows the width of a contour with $\beta=0.5$ ($\log \rho = -4.17$).}
\label{s4GBetaCones}
\end{figure*}

\begin{figure*}
\centering
\includegraphics[width=\linewidth]{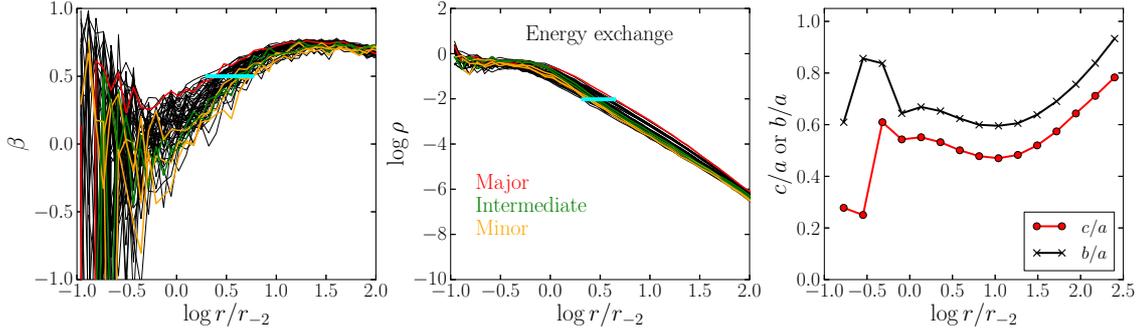}
\caption{This plot shows properties of halo A (a Hernquist profile with $\beta_\text{initial}=0$) from simulation II. As in figure~\ref{s4GBetaCones}, a larger $\beta$ is found along the major axis compared to the minor axis at all radii.}
\label{0EBetaCones}
\end{figure*}

\begin{figure*}
\centering
\includegraphics[width=\linewidth]{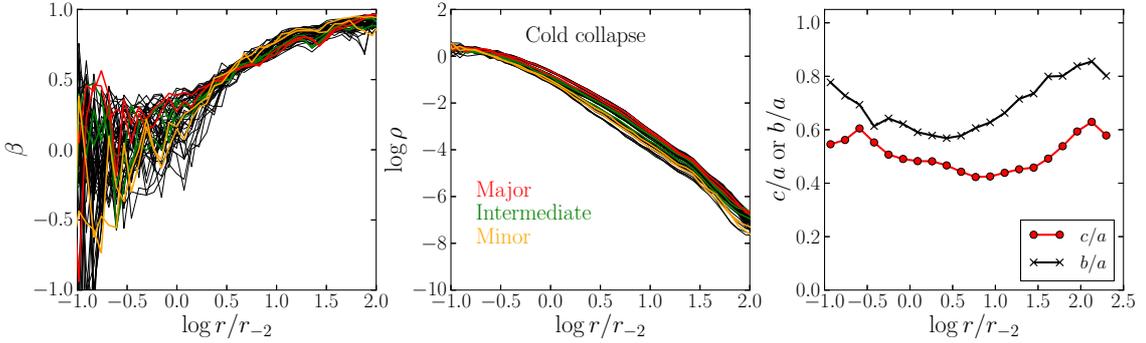}
\caption{Halo I (a Hernquist profile) from the cold collapse simulation. In the inner parts $\beta$ is largest along the major axis, whereas it is larger along the minor axis in the outer parts.}
\label{ColdCollapseBetaCones}
\end{figure*}

\begin{figure*}
\centering
\includegraphics[width=\linewidth]{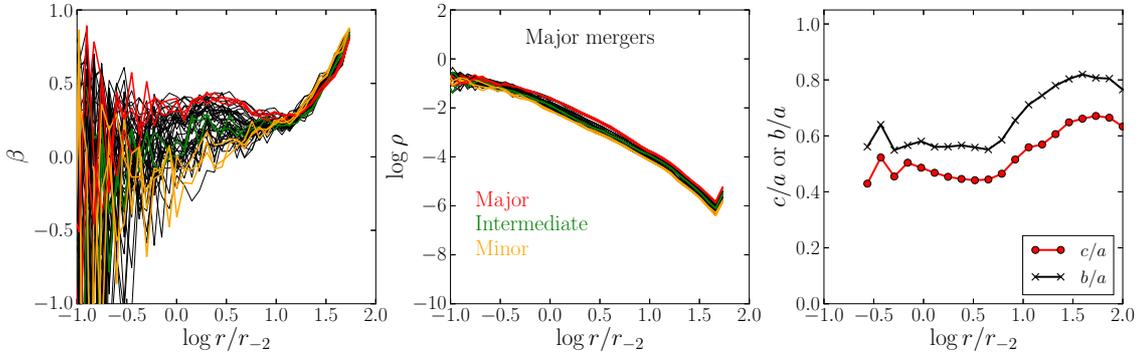}
\caption{Halo K (a Hernquist profile with $\beta_\text{initial}=0$) from the merger simulation. Along the major axis $\beta$ is roughly constant in the inner parts, and along the minor axis it is monotonically increasing.}
\label{MergerCones}
\end{figure*}

\begin{figure*}
\centering
\includegraphics[width=\linewidth]{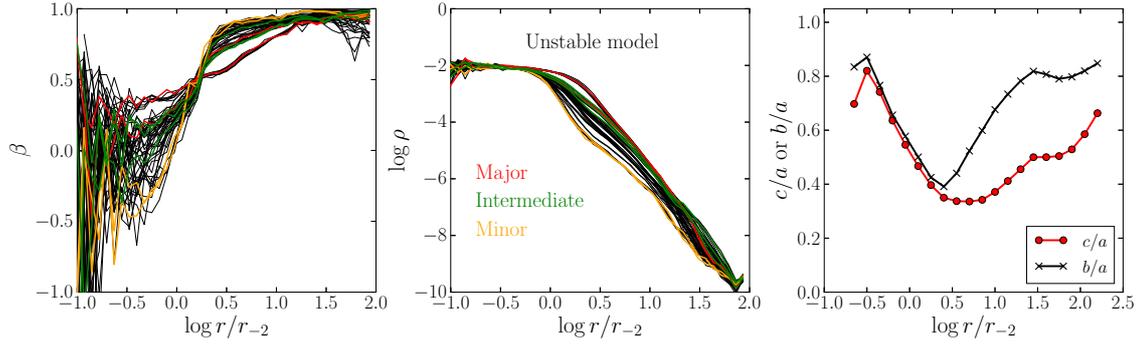}
\caption{Halo R  ($\rho\propto 1/(1+r)^5$) from the instability simulation. The density profile of this halo has several bumps and wiggles along the minor axis. $\beta$ also exhibits different behaviour along different axes.}
\label{ROI2}
\end{figure*}

\begin{figure*}
\centering
\includegraphics[width=\linewidth]{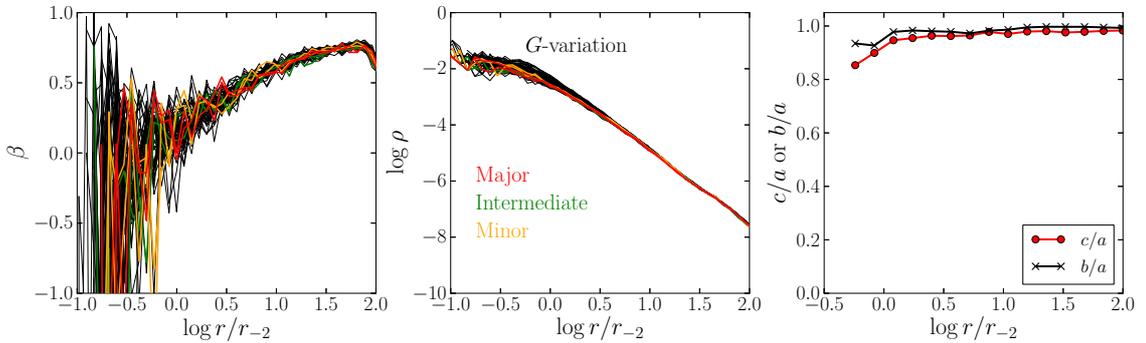}
\caption{Halo A (a Hernquist profile with $\beta_\text{initial}=0$) from the $G$-variation simulation. This halo has a spherical shape and a spherically symmetric velocity anisotropy.}
\label{0GBetaCones}
\end{figure*}

\subsection{A summary of the categories}\label{catg}

To summarise the behaviour of velocity anisotropy and shape in our haloes, we have grouped them into three categories:

\begin{enumerate}
\item One category is the triaxial haloes with $\beta$ differing from cone to cone. The contours of constant $\beta$ are typically more elongated than the constant-$\rho$ contours. Examples of such haloes are shown in figure \ref{s4GBetaCones} and \ref{0EBetaCones}.
\item \label{Type2}We also see examples of triaxial haloes, where the constant-$\beta$ contours differ significantly from triaxial ellipsoids (see figure~\ref{ColdCollapseBetaCones}-\ref{ROI2}). We do for example see haloes, where $\beta$ is largest along one axis in the inner parts, and smallest along this same axis in the outer parts.
\item In our simulations we also see almost spherical structures, where $\beta$ is behaving similarly along each axis (see figure~\ref{0GBetaCones}).
\end{enumerate}

Note, that the above list is not supposed to be a comprehensive list of all possible configurations of $\beta$- and $\rho$-profiles in dark matter haloes. The purpose of the list is to highlight features, which have not been extensively studied in the literature.

\section{Relating the results to cosmological haloes} \label{ch:cosmo}

\subsection{Comparison with a cosmological halo}

A study which also examines halo shapes and the direction-dependence of $\beta$-profiles is the analysis of the Via Lactea II halo, where the probability distribution of the velocity anisotropy, $P(\beta)$, is calculated in four different radial bins, and along the major, intermediate and minor axes (see figure 7 from \citep{2009MNRAS.394..641Z}). The Via Lactea II halo has a behaviour, which in several ways is consistent with our merger simulation in figure~\ref{MergerCones}. In both studies $\beta$ is largest along the major axis ($\beta\simeq 0.3-0.4$) and smallest along the minor axis ($-0.6\lesssim\beta\lesssim 0.0$) in the inner parts. The values of $\beta$ along the intermediate axis are somewhere in between. The halo from the Via Lactea II simulation is close to being prolate ($c/a = 0.52$ and $b/a = 0.62$), which is also similar to the behaviour of our merger remnant. Due to the similarities with our merger remnants, we conclude that the Via Lactea II halo is in category~\ref{Type2} of the categories from section~\ref{catg}.

Merging is not the only process that can produce a halo, which is similar to the Via Lactea II halo. An example is the simulation of the unstable halo (figure~\ref{ROI2}), which has a $\beta$-profile that behaves similarly to the mergers in the inner parts. The unstable halo is, however, significantly more elongated than the Via Lactea II halo.

We encourage future studies of $\beta$-profiles of cosmological haloes to use cones, since it reveals information about halo dynamics, which is hidden when properties are calculated in spherically averaged bins.

\subsection{Spatial anisotropy of galaxy kinematics in galaxy clusters}

In a recent study of galaxy clusters the line of sight velocity dispersion is calculated along the major and minor axis for a stacked sample of 1743 galaxies from the SDSS survey \citep{2012arXiv1208.4598S}. It is found that a significantly larger velocity dispersion is found for galaxies with positions along the major axis. The difference of the velocity dispersions along the two axes is $38\pm13$ km/s. A possible explanation for this difference is the direction-dependence of the velocity anisotropy profile, which we have studied in this article. A different explanation could be that the cluster's non-spherical mass-distribution can produce different line-of-sight velocity dispersions along the minor and the major axis, even though $\beta$ has the same behaviour along these axes.

\begin{figure*}
\centering
\includegraphics[width=\linewidth]{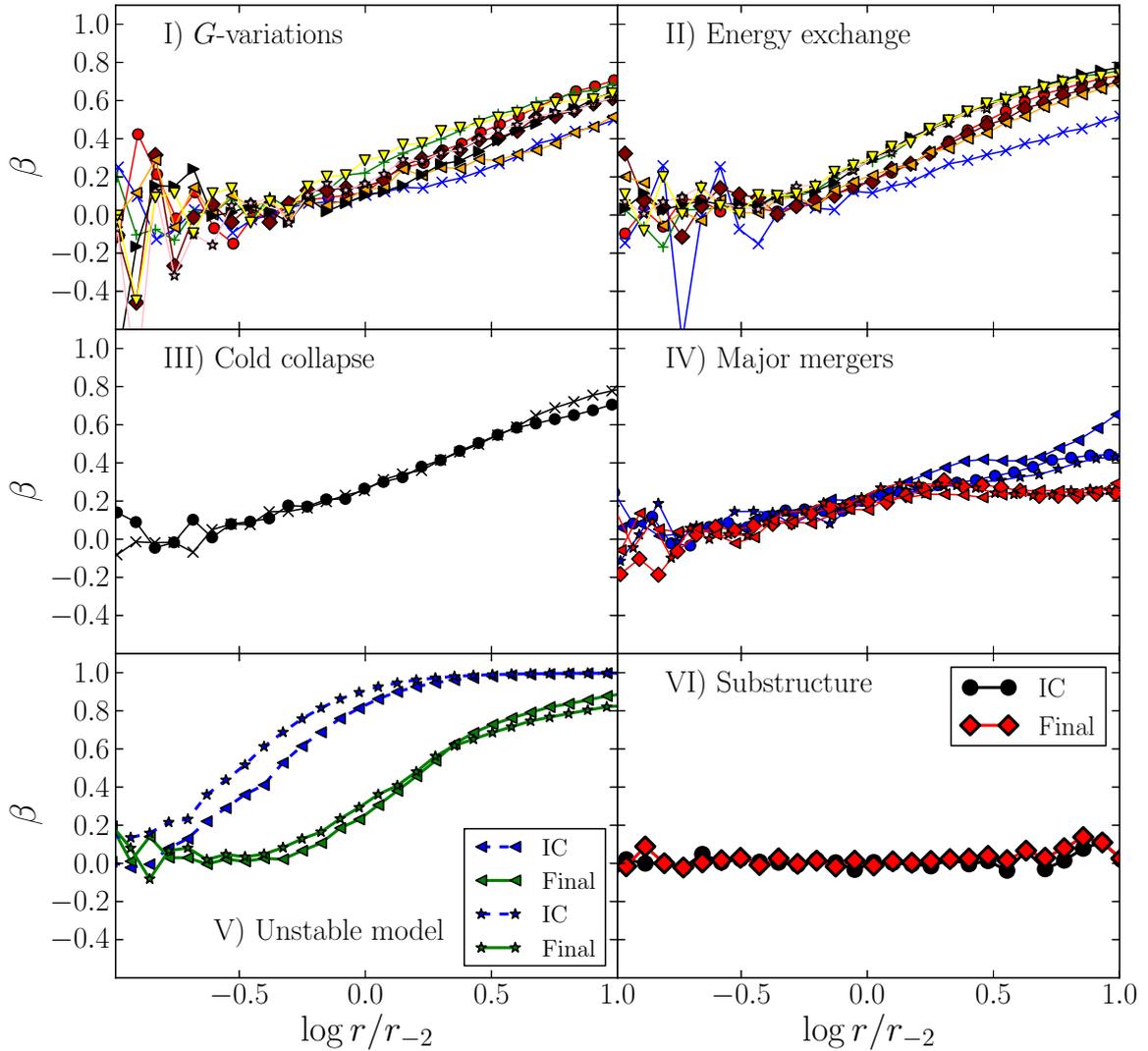}
\caption{The spherically averaged velocity anisotropy profiles for the simulated haloes.}
\label{Fig01:betasph}
\end{figure*}

\begin{figure*}
\centering
\includegraphics[width=\linewidth]{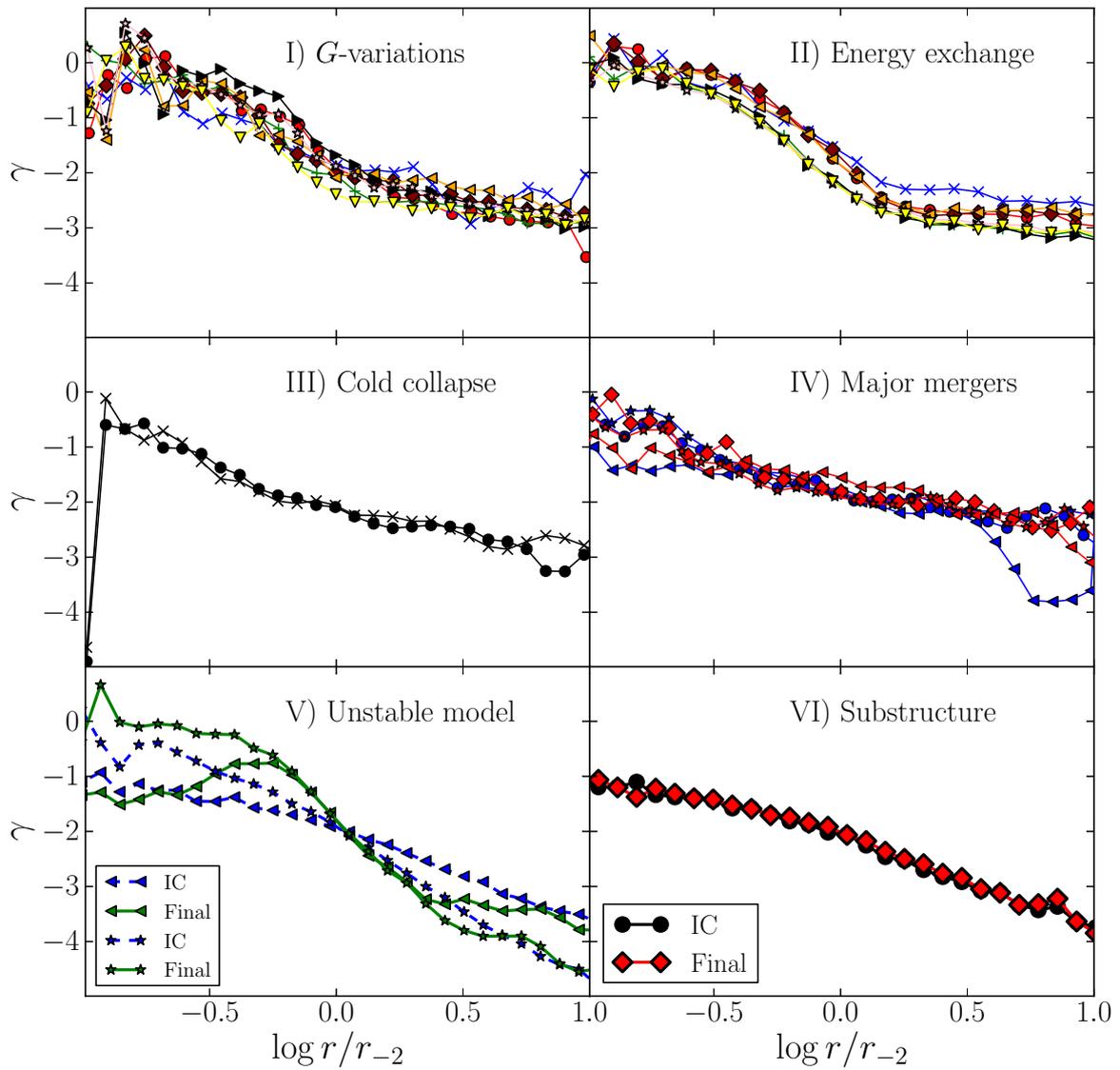}
\caption{$\gamma (r)$ for all the haloes.}
\label{GammaPlot}
\end{figure*}

\begin{figure*}
\centering
\includegraphics[width=\linewidth]{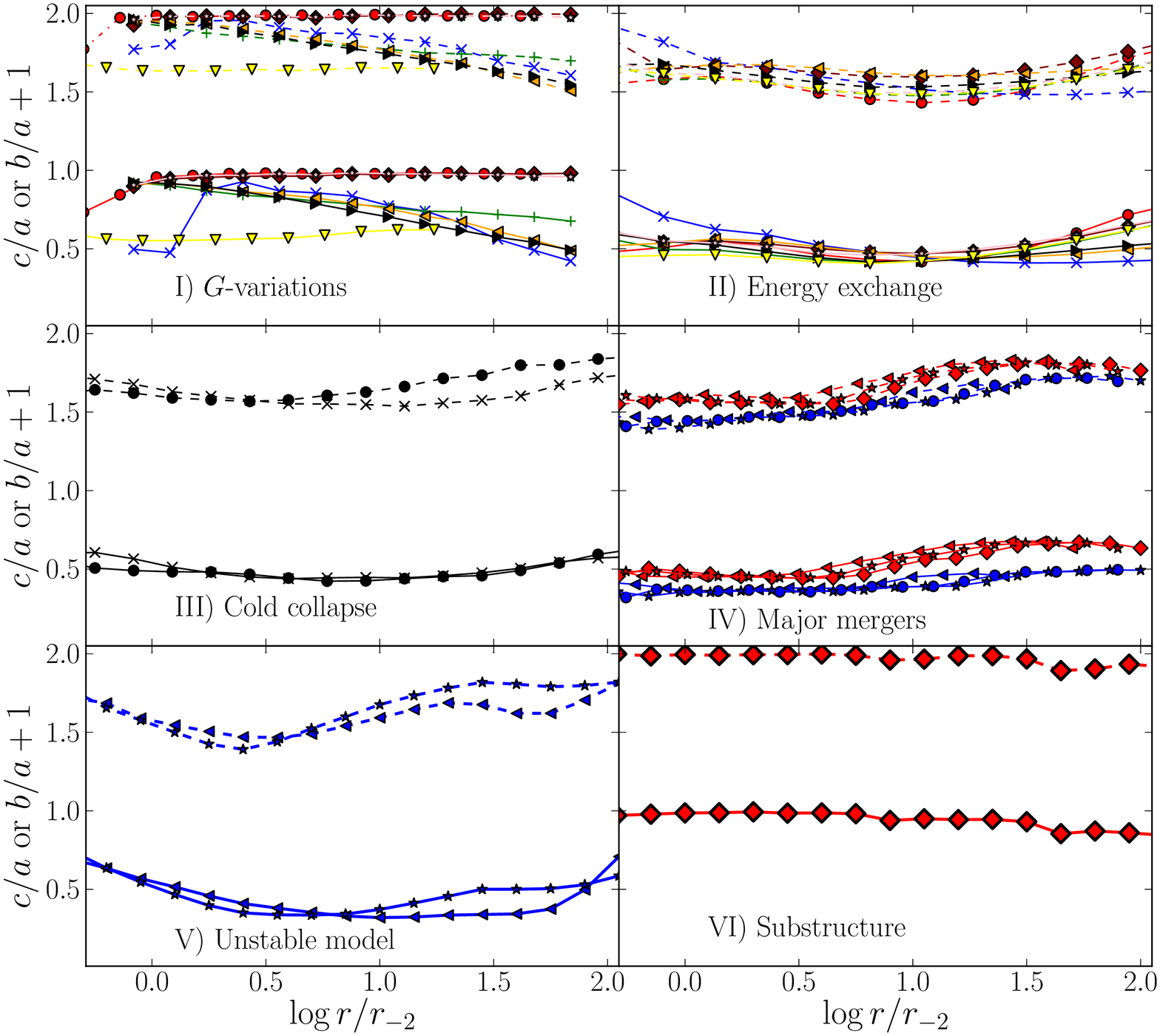}
\caption{The axis ratios, $c/a$ (\emph{solid lines}) and $b/a$ (\emph{dashed lines}), for the haloes. $b/a+1$ is plotted instead of $b/a$ for presentation reasons.}
\label{Shapes02}
\end{figure*}

\section{Spherically averaged profiles}\label{ch:sphaverage}

In this section we will report the spherically averaged profiles of $\beta$ and $\gamma$ together with the halo shapes.

\subsection{Velocity anisotropy profiles}

Different trends are seen in the spherically averaged velocity anisotropy profiles of our haloes, see figure~\ref{Fig01:betasph}. The setups in simulation I, II, III and V produce haloes with $\beta \simeq 0$ in the inner parts and $\beta \simeq 0.8$ in the outer parts. This is consistent with what is e.g. reported in cold collapse simulations \citep{1982MNRAS.201..939V,2006ApJ...653...43M}.

The spherically averaged velocity anisotropy profiles of the merger remnants (simulation IV) are typically increasing from $\beta=0$ in the center of the halo out to the radius with $\beta=0.3$, where a maximum appears. This behaviour is similar to many cosmological haloes \citep{2011MNRAS.tmp..937L}, and in agreement with what is reported in other studies of merger remnants \citep{2004MNRAS.354..522M,2007MNRAS.376.1261M,2012arXiv1205.1799S}.

In the substructure simulation (simulation VI) only a tiny evolution of $\beta (r)$ is seen, which means that the dynamical friction from subhaloes only have a minor effect on velocity anisotropy profiles. This is consistent with the finding that the density profiles of cosmological haloes are not perturbed by the substructure \citep{1999MNRAS.310.1147M}.

\subsection{$\gamma$-profiles}

The six different physical processes in our simulations produce different $\gamma (r)$-profiles, see figure~\ref{GammaPlot}. In simulation I and II a core where $\gamma\simeq0$ appears in the inner regions (with $\log r/ r_{-2} \lesssim -0.5$). Such a core is absent in the remaining simulations. In the outer regions $\gamma (r)$ seems to be around $\gamma \simeq -3$ in simulation I, II and III.

\subsection{Halo shapes}

A huge diversity is seen in the halo shapes in figure~\ref{Shapes02}. Several of the haloes from simulation I (with $G$-variations) are spherical and others are close to being prolate. In simulation II, III, IV and V the shapes of the haloes are close to being prolate. The halo in simulation VI (the simulation with subhaloes) remains spherical.

\clearpage

\section{Comparison with the attractor}\label{sec:attractor}

An aim of the previous sections has been to examine the many different behaviours of halo shapes and $\beta$-profiles of dark matter haloes. In the current section we will instead look for similarities of the different haloes, and compare them with the attractor from HJS.

\subsection{Motivation for the attractor}
The three quantities, $\beta (r)$, $\gamma(r)\equiv d\log \rho / d \log r$ and $\kappa(r)\equiv d\log \sigma_\text{rad}^2 / d \log r$, are clearly of importance for spherical and static distributions of collisionless structures, since the Jeans equation \citep{2008gady.book.....B} can be written as
\begin{align}
-\frac{G M(r)}{r\sigma_\text{rad}^2(r)}=  \gamma(r)+\kappa(r)+2\beta(r).\label{jeans2}
\end{align}
This motivates an analysis of structures behaviour in the $(\beta,\gamma,\kappa)$-space. In the HJS-paper such an analysis was made for the structures from simulation II (the energy exchange simulation), and in this 3-dimensional space all haloes followed a 1-dimensional relation, i.e. an \emph{attractor}. In this section we will compare this attractor prediction with our structures.

\begin{figure*}
\centering
\includegraphics[width=\linewidth]{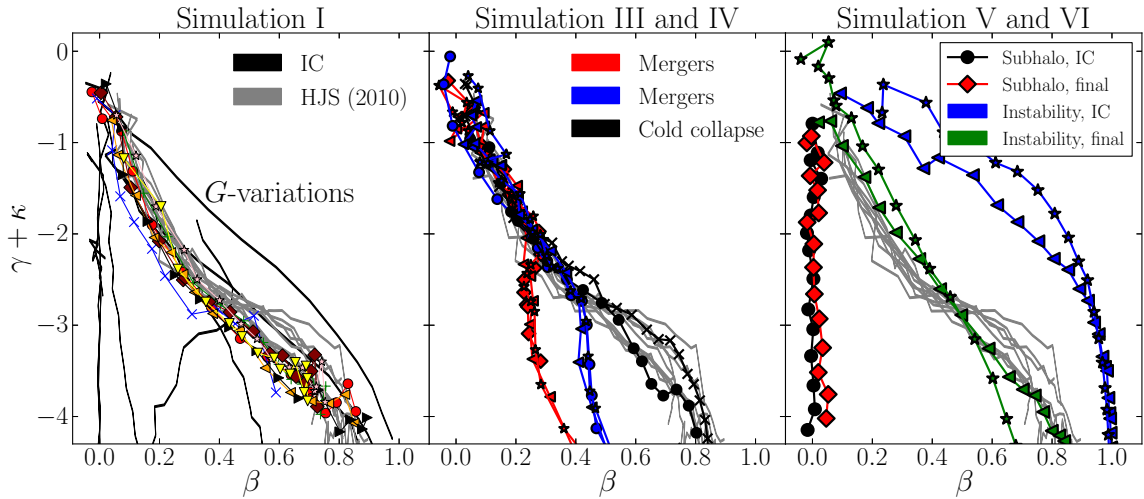}
\caption{Comparison between the attractor from HJS (i.e. simulation II) in \emph{grey}, and the other simulations. We have defined $\kappa\equiv d\log \sigma_\text{rad}^2 / d \log r$ and $\gamma\equiv d\log \rho / d \log r$. In Table~\ref{Table:structures} the different colours and symbols can be found. \emph{Left panel:} The structures exposed to the perturbations, where $G$ was changed instantaneously. \emph{Central panel:} The difference between the \emph{red} lines and the \emph{blue} lines is the orientation of the merger axis in the last simulation (see section~\ref{MajorMergers} for details). Two cold collapse simulations are also shown. \emph{Right panel:} The two haloes from the instability simulation (V) and the subhalo simulation (VI).
}
\label{Fig:attr}
\end{figure*}

\subsection{Comparing haloes with the attractor}

First we will analyse structures in the $(\beta,\gamma+\kappa)$-projection, see figure~\ref{Fig:attr}. The points show properties calculated in spherical bins distributed logarithmically in radius. For simulation I, where $G$ was changed multiple times, the different haloes end up on a one-dimensional curve, like in HJS. The scatter at large $\beta$ is smaller than in the original HJS-simulations, and there is an offset towards smaller $(\gamma+\kappa)$-values in simulation I compared to the HJS-simulations (simulation II). The same trend is seen in the study of the attractor by \citep{2012arXiv1204.2764B}. The two haloes from the cold collapse simulations (simulation III), also end up on the attractor in this projection.

Simulation IV involves multiple major mergers. The result of the three simulations with different density profiles are shown as \emph{red points} in figure~\ref{Fig:attr}. The \emph{blue points} show the additional simulations with identical collision axes in the last two simulations. The spherically averaged profiles follow the attractor in the inner parts, even though $\beta$ behaves different along each axis (see figure~\ref{MergerCones}). When inspected the attractor is not obeyed when $\beta$, $\gamma$ and $\kappa$ are calculated in cones.

The unstable model in Simulation V ends up close to the attractor. The halo in simulation VI is not perturbed enough by the presence of the substructures to be dragged towards the attractor.

\subsubsection{Different projections}

\begin{figure}
\centering
\includegraphics[width=\linewidth]{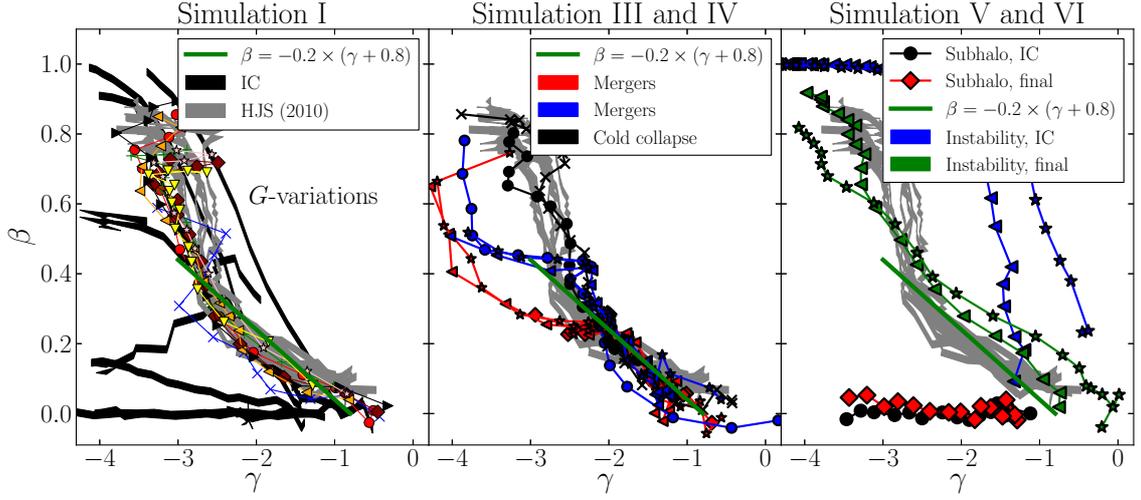}
\caption{A projection in the $(\gamma,\beta)$-plane. Also plotted is the linear relation (\emph{thick green line}) from \cite{2006JCAP...05..014H}.}
\label{gamma_beta}
\end{figure}

\begin{figure}
\centering
\includegraphics[width=\linewidth]{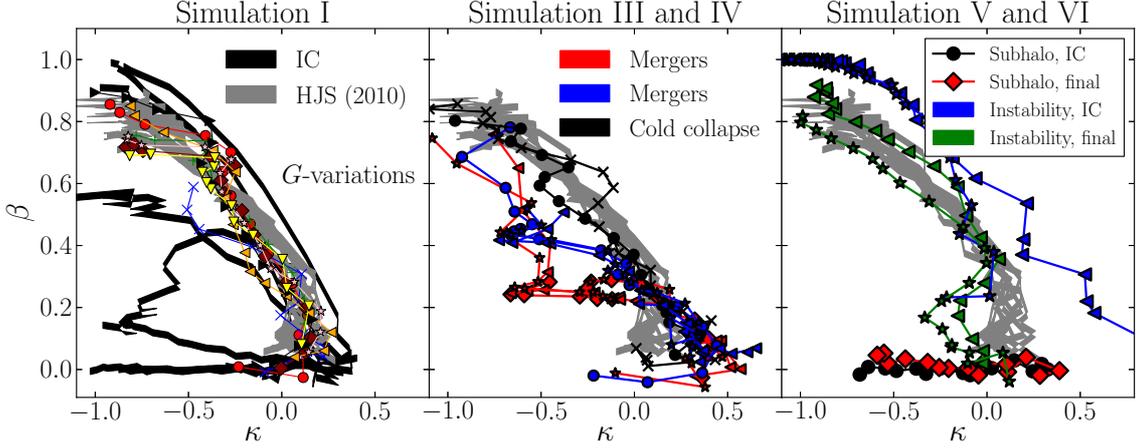}
\caption{The $(\kappa,\beta)$-projection.}
\label{beta_kappa}
\end{figure}

In figure~\ref{gamma_beta}-\ref{gamma_kappa} the structures are plotted in the three principal projections, i.e. ($\gamma,\beta$), ($\kappa,\beta$) and ($\gamma,\kappa$). In all projections there is a very good agreement between simulation I and II, so from now we will define the attractor by the result of these two simulations.

In the $(\gamma,\beta)$-projection (figure~\ref{gamma_beta}) the cold collapse process produces haloes, which are consistent with the attractor prediction. The spherically averaged properties of the merger remnants follow the attractor in the inner parts. We also see that the haloes from simulation I-IV follow the linear $\gamma$-$\beta$ relation from \cite{2006JCAP...05..014H} in the inner parts, where $\gamma \gtrsim -2.2$. The unstable model is not on the attractor.

In the $(\kappa,\beta)$-projection (figure~\ref{beta_kappa}) the collapse simulations are again close to the attractor prediction, but some significant deviations are present in the inner parts with $\beta\lesssim 0.2$. The merger simulations behave differently and the unstable models only follow the attractor (in this projection) in the outer parts with $\beta \gtrsim 0.35$.

\subsubsection{Comparison with cosmological pseudo-phase-space densities}

\begin{figure}
\centering
\includegraphics[width=\linewidth]{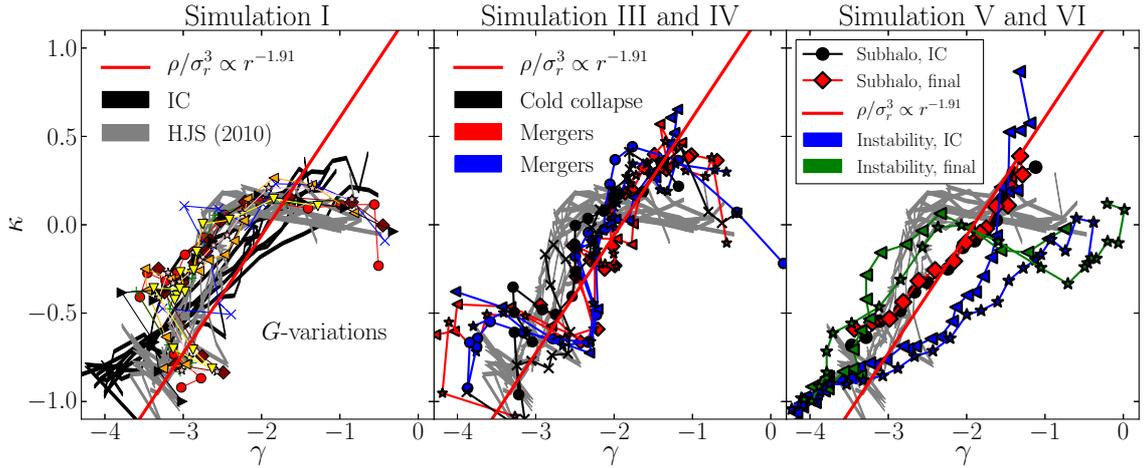}
\caption{The $(\gamma,\kappa)$-projection.}
\label{gamma_kappa}
\end{figure}

Figure~\ref{gamma_kappa} shows the simulations projected onto the $(\gamma,\kappa)$-plane, together with the relation, $\kappa =\frac{2}{3}(\gamma +\alpha)$, which follows from the connection,
\begin{align}
\rho/\sigma^3_\text{rad}(r)\propto r^{-\alpha}, \label{psd}
\end{align}
with $\alpha=1.91$ \citep{2011MNRAS.tmp..937L}, which comes from fitting cosmological haloes over their entire radial ranges.

We see that the attractor prediction is inconsistent with the relation from Eq. \eqref{psd}, which is obeyed by cosmological haloes. It can also bee seen that the inner parts of the merger remnants and the haloes from the collapse simulations are closer to relation \eqref{psd} than the attractor prediction.

\begin{figure}
\centering
\includegraphics[width=0.7\linewidth]{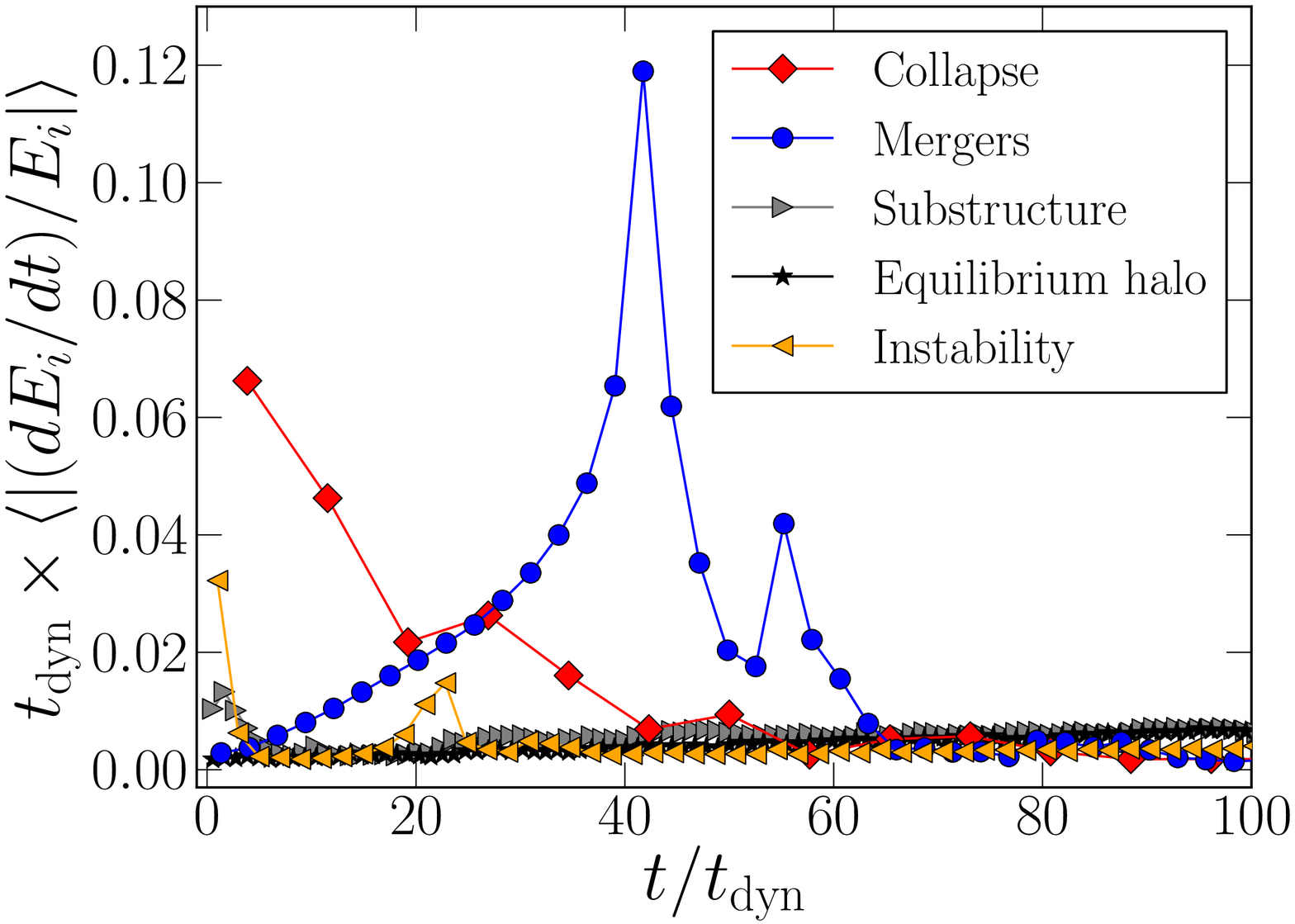}
\caption{The fractional change in energy per dynamical time for particles that end up in a spherical shell around $r_{-2}$ in the final snapshots. In the cold collapse simulation and the merger simulations large changes in the energy of the particles appear. Much lower variations in the energy are present in the simulation with the substructure. For reference an equilibrium structure with no substructure is also shown.}
\label{Fig:dE}
\end{figure}

\subsubsection{Energy exchange in the relaxation processes}\label{relax-process}

The purpose of this subsection is to study the change of particle energies in the relaxation processes. For our structures we identified the particles, that ended up in a thin spherical shell centered at $r_{-2}\equiv r(\gamma=-2)$ in the last snapshot, and monitored how their energies changed through the simulation. In figure~\ref{Fig:dE} we have plotted the dimensionless quantity $t_\text{dyn}(r_{-2})\times \langle |(dE_i/dt)/E_i|\rangle$ ($E_i$ is the energy of the $i$'th particle, and $\langle \cdots \rangle$ denotes the mean of the particles that ended up in the shell at $r_{-2}$) as function of $t/t_\text{dyn}(r_{-2})$.

For the cold collapse experiment (the structure following a Hernquist profile is shown), the collapse begins immediately after the start of the simulation, and a redistribution of the energy occurs. For the merger experiment (the Hernquist profile is shown) the exchange of energy grows in the beginning, where the two haloes approach each other, and the largest peak occurs approximately when the two haloes collide for the first time ($t\simeq 40t_\text{dyn}$), and a smaller peak is present when they collide for the second time ($t\simeq 55t_\text{dyn}$). At later times the energy exchange is more than an order of magnitude lower than the peak values for both experiments.

The energy exchange in the simulation with an unstable model (the structure following Eq.~\eqref{rho05} is shown) is weaker than in the collapse simulation and the mergers. The instability causes the energy-exchange around the peak at $t=22t_\text{dyn}$. When inspecting the $\beta$-profile in each individual snapshot it is seen, that it changes shape from $t/t_\text{dyn}=16$ until $t/t_\text{dyn}=24$. In the later snapshots no significant evolution of this halo is seen.

In the subhalo simulation a small energy exchange is present throughout the simulation. To see the amount of energy exchange generated by the substructure, we ran a similar simulation with a halo in equilibrium; i.e. a halo where the subhaloes were removed, and ordinary particles were distributed throughout the halo to compensate for the removed mass. By comparing the two simulations we find that only the small bump at $t\lesssim 5t_\text{dyn}$, where some of the particles in the substructure are stripped, might be caused by the subhaloes.

\subsection{An overview: which structures are on the attractor?}

The overall pattern in our simulations is that the haloes that end exactly on an attractor are those from simulation I and II, which involved artificial processes that changed the gravitational constant or the kinetic energy of the particles. The collapse simulations ended up close to the attractor, but not exactly on it, and the merger simulations had clear deviations in the outer parts.

We also found that haloes following the attractor deviate from the universality, $\rho/\sigma^3_\text{rad}\propto r^{-\alpha}$, which is found in cosmological simulations. Our conclusion is that the dynamics and structure of cosmological haloes are different from the attractor. Further indications, that cosmological haloes are not on the attractor, comes from the finding that our cold collapse and merger simulations, which are both cosmologically realistic processes, produce haloes that are not exactly on the attractor. The strong direction-dependence of $\beta$ found in the Via Lactea II halo \citep{2009MNRAS.394..641Z} also supports that cosmological haloes are not on the attractor.

For both the merger simulations and the collapse simulations we do, however, see that the $\beta$- and $\gamma$-profiles (spherically averaged) in the inner parts of the structures obey the attractor in the $(\gamma,\beta)$-projection. Since a very good agreement only is found in this projection, cosmological haloes are better described by the $\gamma$-$\beta$ relation from \citep{2006NewA...11..333H} than by the attractor.

It is important to note that this $\gamma$-$\beta$ relation only applies to the spherically averaged values of $\beta$- and $\gamma$-profiles. The haloes still have freedom to have different $\beta$-profiles in different directions as long as the spherically averaged profiles obey this relation. An example of a halo in which the $\beta$-$\gamma$ relation is not obeyed through cones in different directions, but only in spherically averaged bins, has been presented in the study of merger remnants in \citep{2012arXiv1205.1799S}.

\subsection{The relevance of the attractor}

We have shown that typical cosmological haloes have departures from the attractor prediction. Even though this is the case, we still believe that there must be a physical origin of the similar behaviour of the haloes from simulation I and II, when they are analysed in the $(\beta,\gamma,\kappa)$-space. Understanding this behaviour could potentially lead to a better understanding of violent relaxation and mixing processes in collisionless systems.

\section{Summary}

In this article we have examined velocity anisotropy profiles and halo shapes in a range of pure dark matter simulations. We have found that haloes with elongated shapes can have several kinds of direction-dependent velocity anisotropies. In some cases the $\beta$-profile is aligned with the density profile, and the largest velocity anisotropy is found along the major axis. In other cases the $\beta$-profiles are more complicated, and it is not possible to define an axis along which the velocity anisotropy is largest at all radii. Such a behaviour is e.g. seen in the remnant of a major merger, which has a shape and a direction-dependence of $\beta$, that is very similar to the Via Lactea II halo. We suggest that future studies of cosmological haloes should calculate the velocity anisotropy profiles in cones, since it reveals properties of the true velocity distributions, which are hidden from the spherically averaged profile.

We have also compared our haloes with an attractor \citep{2010ApJ...718L..68H}, and we conclude that cosmological haloes have departures from the attractor prediction. We do, however, find that the spherically averaged $\beta$- and the $\gamma$-profile ($\gamma$ is the logarithmic derivative of the density profile) obey an approximately linear relation (from \citep{2006NewA...11..333H}) in the inner parts, even though this relation is not necessarily obeyed for particles in individual cones pointing in different directions.

\acknowledgments It is a pleasure to thank Jens Hjorth for discussions and comments. We also thank for useful referee comments. The simulations were performed on the facilities provided by the Danish Center for Scientific Computing. The Dark Cosmology Centre is funded by the Danish National Research Foundation.\\

\def\aj{AJ}
\def\araa{ARA\&A}
\def\apj{ApJ}
\def\apjl{ApJ}
\def\apjs{ApJS}
\def\apss{Ap\&SS}
\def\aap{A\&A}
\def\aapr{A\&A~Rev.}
\def\aaps{A\&AS}
\def\mnras{MNRAS}
\def\na{New Astronomy}
\def\jcap{JCAP}
\def\nat{Nature}
\def\pasp{PASP}
\def\aplett{Astrophys.~Lett.}
\def\physrep{Physical Reviews}
\bibliographystyle{JHEP}

\providecommand{\href}[2]{#2}\begingroup\raggedright\endgroup

\end{document}